\documentclass[11pt,aps,onecolumn,superscriptaddress]{revtex4-2}

\usepackage[T1]{fontenc}
\usepackage[utf8]{inputenc}
\usepackage[english]{babel}
\usepackage{graphicx}

\makeatletter
\renewcommand{\p@subsection}{}
\renewcommand{\p@subsubsection}{}
\makeatother

\usepackage[hyperfootnotes=false]{hyperref}
\usepackage{amsfonts, amsbsy, amssymb, amsmath}

\begin{document}
\title{Predicting trajectory behaviour via machine-learned invariant manifolds}
 \author{Vladim{\'i}r Kraj{\v{n}}{\'a}k}
 \email{v.krajnak@bristol.ac.uk}
 \affiliation{School of Mathematics, University of Bristol, Fry building, Woodland Road, Bristol BS8 1UG, United Kingdom}
 \author{Shibabrat Naik}
 \email{shibabratnaik@gmail.com}
 \affiliation{School of Mathematics, University of Bristol, Fry building, Woodland Road, Bristol BS8 1UG, United Kingdom}
 \author{Stephen Wiggins}
 \email{s.wiggins@bristol.ac.uk}
 \affiliation{School of Mathematics, University of Bristol, Fry building, Woodland Road, Bristol BS8 1UG, United Kingdom}

\begin{abstract}
In this paper, we use support vector machines (SVM) to develop a machine learning framework to discover phase space structures that distinguish between distinct reaction pathways. The SVM model is trained using data from trajectories of Hamilton's equations and works well even with relatively few trajectories. Moreover, this framework is specifically designed to require minimal a priori knowledge of the dynamics in a system. This makes our approach computationally better suited than existing methods for high-dimensional systems and systems where integrating trajectories is expensive. We benchmark our approach on Chesnavich's CH$_4^+$ Hamiltonian.
\end{abstract}

\maketitle

\section{Introduction}

Recent breakthroughs in computer vision, natural language processing, self-driving vehicles, medical diagnostics, brain-computer interfaces, robotics, particle physics, nano sciences spurred on by data-driven methods such as machine learning (ML), deep learning (DL), and reinforcement learning (RL) has sparked the interest in leveraging these methods’ capabilities for approximation and extrapolation in computational and theoretical chemistry~\cite{rupp_guest_2018,Cova2019,unke_high-dimensional_2020,Meuwly2021,westermayr_perspective_2021, Ceriotti2021}. In the context of chemical physics, the main focus has been on learning force field parameters from ab initio calculations~\cite{li_machine_2017,unke_high-dimensional_2020,qiao_orbnet_2020,Noe2020}, building accurate potential energy surfaces~\cite{koner_permutationally_2020, Qu2021} and identifying low-dimensional geometries of high-dimensional systems \cite{Bittracher2017,Bittracher2020}. Other works focus on constructing dividing surfaces using the dynamical systems viewpoint~\cite{pozun2012optimizing,schraft_neural_2018} and accurately classifying or predicting the behaviour of trajectories \cite{carpenter_empirical_2018,naik_support_2021,Maley2021}. In this paper we use support vector machines (SVM) to develop a machine learning framework to discover the phase space structures that can distinguish between distinct reaction pathways. The machine learning model is trained using trajectory data from Hamilton's equations but lends itself for use in molecular dynamics simulation. The framework is specifically designed to require minimal a priori knowledge of the dynamics in a system. We benchmark our approach with a model Hamiltonian for the reaction of an ion and a molecule due to Chesnavich \cite{Chesnavich1986}. In particular, Chesnavich considers the reaction 
\begin{equation}
\text{CH}_4^{+} \rightleftharpoons \text{CH}_3^{+} + \text{H},
\end{equation}
\noindent
and a model consisting of two parts: a rigid, symmetric top representing the $\text{CH}_3^{+}$ ion, and a mobile $\text{H}$ atom. A detailed derivation of this model can be found in reference \cite{ezra2019chesnavich}. The reaction of an ion and a molecule is a topic of fundamental  and continuing interest  in chemistry, and the topic is reviewed in \cite{futrell1968ion, stevenson1957ion, ferguson1975ion, Chesnavich82, franklin2012ion, meyer2017ion}.

In recent years this model has seen applications to understanding mechanisms responsible for the roaming reaction mechanism \cite{Bowman2011Suits}. Roaming is a new mechanism for dissociation of a molecule. In $\text{CH}_4^{+}$ it corresponds to the  dissociation of a hydrogen atom.  In \cite{mauguiere2014multiple} it is asserted that a dissociating molecule should possess two essential characteristics in order to label the reaction as ``roaming''. In particular, the molecule should have competing dissociation channels, such as dissociation to molecular and radical products, and there should exist a long range attraction between fragments of the molecule. In a recent review article \cite{suits2020roaming} Suits refines this definition of roaming even further by stating that  ``A roaming reaction is one that yields products via reorientational motion in the long-range ($3-8$\AA) region of the potential.'' Thus, a method that can distinguish between roaming, direct dissociation, and other competing reaction pathways is of interest in understanding mechanisms.

In \cite{ mauguiere2014multiple, mauguiere2014roaming, mauguiere2017roaming, krajnak2018phase} phase space analyses of roaming were carried out that first determined  invariant geometrical structures in phase space that formed the boundaries between the competing dissociation channels. In other words, the strategy was to determine the phase space structures that separated the trajectories into different classes of dynamical behavior. Our machine learning approach is different, and effectively an opposite approach. We start with trajectories (``the data'') and use support vector machines to determine the boundaries between initial conditions corresponding to different classes of trajectories. While it would appear that this is a more direct approach in the sense that we are dealing first, and directly, with trajectories, it will turn out that these boundaries between different classes of trajectories are actually invariant phase space structures of the same type observed in the earlier analyses of Chesnavich's model.

Our goal is to train a machine learning model using a labeled data set consisting of initial conditions on a two dimensional section of the energy surface. The labels denote the distinct end products formed via distinct reaction pathways. We use support vector machines (SVM), a class of supervised learning classification and regression algorithms~\cite{Cortes95, Vapnik96,Vapnik2000}, which has been used for chaotic time series regression in~\cite{Mukherjee1997} and construction of dividing surfaces \cite{pozun2012optimizing}. We employ the classification algorithm, which constructs a decision function with maximum margin to approximate the boundary between different classes of data. In our case, the classes in the classification correspond to initial conditions leading to ``qualitatively different'' dynamical behaviour. The learned boundary between trajectories leading to distinct end product is also verified using the dynamical systems theory of reactive islands~\cite{DeVogelaere1955,PollakChild80,berne_isomerization_1982,de_leon_order_1989,marston_reactive_1989,OzoriodeAlmeida90,naik_support_2021} and therefore we confirm that the SVM approach enables us to determine the geometrical structures in phase space governing distinct reaction pathways.

We choose support vector machines for the following reasons. (i) We aim to approximate nonlinear boundaries between classes of trajectories, for which nonlinear kernels in SVM are designed. (ii) SVM has been known to provide good predictions even with small training datasets.  Consequently our approach is well-suited for systems where obtaining training data is expensive either due to computational complexity or high dimension. This makes our SVM-based approach a computationally efficient gateway to the approximation of phase space structures in higher dimensional systems, such as the system-bath models of isomerization~\cite{naik2020}. It also offers the possibility to capitalise on the recent extension of reactive island theory to systems with three DoF~\cite{krajnak2021reactive}.

We would like to remark on what we mean by minimal a priori knowledge of the dynamics. In contrast to existing methods for discovering reactive trajectories, our approach does not require any information about phase space structures, such as normally hyperbolic invariant manifolds and associated stable/unstable invariant manifolds, phase space dividing surface, or configuration space structure such as minimum energy paths. Our approach circumvents the need for a complete dynamical systems analysis and potentially costly numerical calculations to obtain such structures. The approach needs as input two or more well-defined (based on chemical intuition of product formation) classes of trajectories or a good approximation thereof, and a surface to launch initial conditions. If available, additional information may be built into the approach, but for benchmarking purposes we present our approach using only trajectory data.

This paper is outlined as follows. In Section \ref{sec:Ches} we describe the Chesnavich model and some of its previous uses. In Section \ref{sec:sampling} we describe a data efficient procedure to sample initial conditions of trajectories based on the idea of active learning. In Section \ref{sec:classes} we describe four different classes on trajectories based on expected behavior in the system. In Section \ref{sec:DT} and Section \ref{sec:AT} we provide a more detailed discussion of how we define the trajectory classes of dissociation and association based on the properties and geometry of the potential energy surface. In Section \ref{sec:surfaces} we discuss the choice of initial surfaces from which we sample initial conditions  requiring minimal a priori knowledge of the dynamics in the system that guide our initial classification of trajectories. In Section \ref{sec:active} we discuss the active learning of trajectory classes without the need to compute trajectories. In Section \ref{sec:structures} we discuss the phase space structures that divide classes of trajectories known from previous work. In Section \ref{sec:comparison} we discuss how our machine learning approach based on SVM recovers these phase space structures. In Section \ref{sec:concl} we describe our conclusions and some possible future directions of investigation. Supplementary information contains further details about SVM and discusses some issues related to how SVM deals with the type of fractal structures arising from the intersection of stable and unstable manifolds.

\section{Model and method}

\subsection{Chesnavich's CH$_4^+$ system}
\label{sec:Ches}

Chesnavich's CH$_4^+$ system is a phenomenological model with 2 degrees of freedom introduced in \cite{Chesnavich1986,ezra2019chesnavich} to study CH$_4^+$ $\rightleftharpoons$ CH$_3^+$ $+$ H, representing reactions that involve the passage through multiple transition states. The system was also investigated in the context of roaming \cite{mauguiere2014multiple,mauguiere2014roaming,krajnak2018phase,krajnak2018influence}.

Chesnavich's CH$_4^+$ system is defined by the Hamiltonian
\begin{equation}
H(r,\theta,p_r, p_\theta) = \frac{1}{2} \frac{p_r^2}{\mu} + \frac{1}{2}p_\theta^2 \left(\frac{1}{\mu r^2}+\frac{1}{I_{CH_3}}\right) + U(r,\theta),
\label{eq:chesHam}
\end{equation}
where $(r,\theta)$ are the polar coordinates describing the position of the mobile H atom with respect to the centre of mass of the rigid body CH$_3^+$ and $p_r, p_\theta$ are the respective canonically conjugate momenta. The parameter $\mu$ is the reduced mass of the system $\mu=\frac{m_{CH_3}m_{H}}{m_{CH_3}+m_{H}}$, where $m_{H}=1.007825$ u and $m_{CH_3}=3m_{H}+12.0$ u, and $I_{CH_3}=2.373409$~u\AA$^2$ is the  moment of inertia of CH$_3^+$. A contour plot of $U$ is shown in Fig. \ref{fig:types}. For simplicity we consider the system at a fixed total energy $E=2$ kcal/mol. We remark that we obtain similar results for $E=1$ kcal/mol and the system does not exhibit roaming for $E \geq 2.5$ kcal/mol \cite{krajnak2018phase}.

\begin{figure}
\centering
\includegraphics[width=0.4\textwidth]{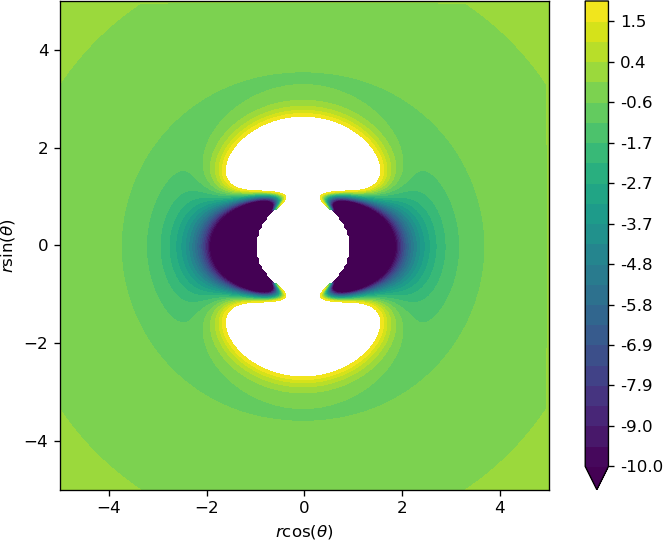}
\caption{A contour plot of potential energy surface $U$ accessible at total energy $E=2$ kcal/mol. Potential wells are shown in shades of blue, while the white area is inaccessible the this energy.}
\label{fig:pot}
\end{figure}

In Fig. \ref{fig:pot}, we show a contour plot of the potential\cite{Chesnavich1986} $U(r,\theta)$ which is characterised by two potential wells separated by two areas of high potential at a short distance away from the origin. With increasing $r$, the potential becomes increasingly independent of $\theta$ and converges to zero from below, $U\approx -r^{-4}$. 

Since the system is unbounded for $E>0$, the ergodic hypothesis fails and Poincar\'e recurrence does not apply at the energies considered here.

\subsubsection{Sampling procedure to generate training data}
\label{sec:sampling}

Sampling the energy surface using a regular grid on a two-dimensional surface involves a considerable amount of data from regions with dynamics that is fairly regular on a short timescale. Furthermore, the accuracy of the learned decision boundary is also limited by the amount of training data and its spacing, and thus the learning will worsen as the volume of the energy surface increases. A significantly less data-intensive approach is offered by active learning \cite{settles09,Kremer14}, where the ``learner'' biases its sampling based on information obtained from previous samples.  Following the principles of active learning \cite{settles09,Kremer14}, the work of  \cite{pozun2012optimizing} and our recent work on H\'enon-Heiles Hamiltonian~\cite{naik_support_2021}, it is more data-efficient to add sample points gradually based on the coarse boundary learned from the previous step. Briefly, we initialize training a support vector classifier with a coarse grid of data points and iteratively add data points on the decision boundary learned from the previous step and re-run SVM with the expanded training data. In this work we start with a coarse regular grid (5x5 or 10x10) and subsequently introduce additional points randomly selected using a uniform distribution on the boundary between classes as predicted by SVM. This allows the algorithm to explore the intricate structures (due to ``twists'' and ``turns'' in the phase space) usually formed by the globalized invariant manifolds that mediate the evolution of a trajectory.

We would like to point out the importance of the precise problem formulation. Homoclinic and heteroclinic intersections of invariant manifolds lead to fractal structures, that is a fractal boundary between classes of dynamics. There is no known way to resolve fractal structures with finite precision and a finite number of data points. Thus approximating the boundary using fixed-width Gaussians is bound to fail. In many systems it is possible to avoid these fractal structures by carefully selecting a surface of initial conditions and studying the dynamics under the corresponding (Poincar\'e) return map.

\subsection{Definition of trajectory classes}
\label{sec:classes}

\begin{figure}
 \centering
 \includegraphics[width=0.48\textwidth]{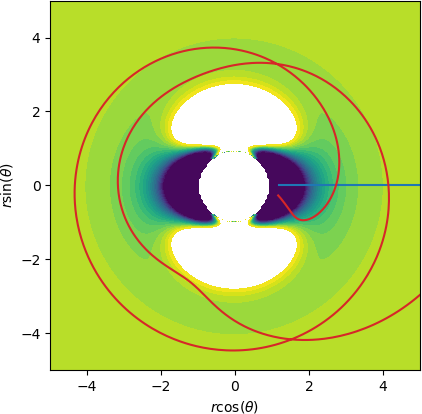}
 \includegraphics[width=0.48\textwidth]{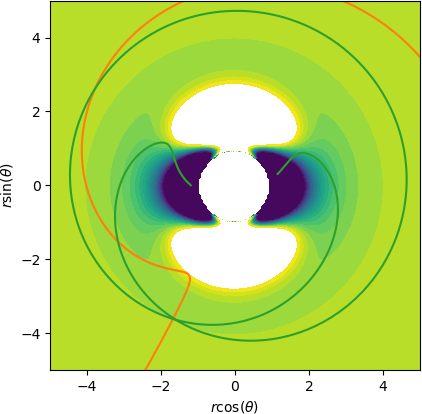}
 \caption{Representative trajectories exhibiting direct dissociation (blue), roaming (red), isomerisation (green) and non-reactivity (yellow) plotted on top of a contour plot of potential energy $U$. Intersections of these trajectories with the middle DS are indicated in Fig. \ref{fig:islands}.}
 \label{fig:types}
\end{figure}

Trajectories in this system exhibit dynamical behaviour of four types: direct dissociation, isomerisation, non-reaction and roaming. Examples of trajectories of each type are shown in Fig. \ref{fig:types}. These trajectories behave differently to each other, because they are separated by phase space structures asymptotic to unstable periodic orbits that govern the dynamics in this system. A precise definition is based on dividing surfaces constructed using unstable periodic orbits, as dissociation and access to the well are controlled by unstable periodic orbits \cite{mauguiere2014roaming}, a short summary can be found in Supplementary information. The boundaries between the four types of trajectories are known to be fractal \cite{krajnak2018phase}.

To overcome the difficulties posed by fractal boundaries, we focus on learning and predicting {\em imminent dissociation} and {\em imminent association} of CH$_3^+$ and H. Given a two-dimensional surface of initial conditions on the energy surface $H(r,\theta,p_r, p_\theta)=E$, {\em imminent dissociation/association} is exhibited by those trajectories that lead to dissociation/association without returning to the initial surface. Based on these properties we distinguish the following classes of trajectories:

\begin{enumerate}
 \item imminent dissociation and association - trajectories that enter a well in backward time and dissociate in forward time,
 \item imminent association only - trajectories that enter a well in backward time and return to the initial surface(s) in forward time,
 \item imminent dissociation only - trajectories that return to the initial surface(s) in backward time and dissociate in forward time,
 \item other - trajectories that return to the initial surface(s) in backward and forward time.
\end{enumerate}

\subsubsection{Dissociation threshold}
\label{sec:DT}

The threshold for considering a trajectory dissociated is of dynamical origin. It is a normally hyperbolic invariant manifold (NHIM) due to a centrifugal barrier, or specifically an unstable periodic orbit in systems with 2 degrees of freedom. It was proven to be present in all systems in which the potential $U \in o(r^{-2})$ as $r\rightarrow\infty$ \cite{krajnak2018phase}. Since the periodic orbit is unstable and rotationally symmetric (required by the proof), all initial conditions with a larger $r$ and $p_r=0$ diverge to infinity. Using the constructive proof, it can be found as a rotationally invariant solution of the system \eqref{eq:chesHam}, given an energy value $E$. In this case a solution is rotationally invariant if $p_r=0$ and
\begin{equation}
 \dot{p}_r = \frac{p_\theta^2}{\mu r^3} - \frac{\partial U}{\partial r}(r,0) = 0,
 \label{eq:prdot}
\end{equation}
where, without the loss of generality, we take the derivative at $\theta=0$.
Following \eqref{eq:prdot} and the energy condition
\begin{equation}
H(r,\theta=0,p_r=0, p_\theta) = \frac{1}{2}p_\theta^2 \left(\frac{1}{\mu r^2}+\frac{1}{I_{CH_3}}\right) + U(r,0)=E,
\end{equation}
we find, in agreement with existing results, that for $E=2$, $\dot{p}_r$ vanishes slightly below $r=7$ and for $E=1$ slightly below $r=10$. We use these values as the threshold for dissociation and remark that the same argument holds regardless of dimension.

\subsubsection{Association threshold}
\label{sec:AT}

Access to each potential well is also governed by a NHIM, that is responsible for allowing some trajectories while repelling others. This creates a dynamical boundary of the potential well and indicates (in phase space) of whether a bond is formed (entry into the potential well) or broken (escape from the potential well). No general way of finding such NHIMs is known, therefore we identify a suitable association threshold indirectly. We use the fact that escape from potential wells in Chesnavich's system is predominantly captured by the dynamics in the radial degree of freedom $r, p_r$ and look at turning points using the Hamiltonian vector field exclusively.

At $E=2$, the vector field at every turning point in the radial direction $\dot{r}=\frac{1}{\mu}p_r=0$ for $1.2\leq r \leq 1.96$ and for every $\theta$ has
\begin{equation}
 \dot{p}_r = \frac{p_\theta^2}{\mu r^3} - \frac{\partial U}{\partial r}(r,\theta) < 0,
\end{equation}
where $p_\theta$ is given implicitly by $H(r,\theta,p_r,p_{\theta})=E$. Therefore all trajectories that are not monotonic in $r$ evolve towards the bottom of the well at $r=1.1$. The same is true for $E=1$ and $1.2\leq r \leq 2.02$. Hence we consider any trajectory that reaches $r=1.9$ as being in the well at these energies.

\subsection{Definition of initial surfaces}
\label{sec:surfaces}

Following our goal to devise an approach that uses minimal a priori knowledge  of the dynamics in the system, we do not fix a surface of section suited to investigate boundaries between the aforementioned classes. Instead, our approach searches multiple surfaces at the same time and identifies the surface on which SVM attains the highest accuracy. Since there may be features that significantly distort the boundaries between trajectory classes or even cause them to be discontinuous, SVM is going to attain the highest accuracy on a surface on which the boundaries are the simplest. As a measure of simplicity, we show the number of support vectors in the accuracy plots in Sec. \ref{sec:active}.

We consider surfaces of the form $r=const$, $\dot{r}>0$ evenly distributed between the association and dissociation threshold from Sec. \ref{sec:classes}. Considering multiple surfaces comes at a low computational costs, as every time we initiate a point on one of the surfaces, we find the intersections (if they exist) with all remaining surfaces during the forward and backward integration that is needed to classify the trajectory. We integrate each trajectory from its initial condition forward and backward in time until it reaches the association threshold, the dissociation threshold, the initial surface(s) or 100 time units.

Before training SVM, we split the data set into $80\%$ training data and $20\%$ test data and calculate accuracy (ratio of correct predictions to total number of predictions), such as in Fig. \ref{fig:adaptive_accuracy}, using unseen data. We remark that most data points lie near the class boundaries in the active learning approach. The calculated accuracy is therefore not comparable to the probability of correctly predicting a random point in the entire domain.

We remark that the choice of surfaces influences the definition of classes, as we terminate trajectories that return to any of the initial $r=const$, regardless of $\dot{r}$. While the surfaces can be defined differently, classification based on imminent dissociation and association cannot be directly related to direct dissociation, roaming and isomerisation without the knowledge of invariant phase space structures. While the majority of the trajectories in each classes can be related to a dynamical behaviour, trajectories near the boundaries are more difficult. As presented in Sec. \ref{sec:comparison}, we still find good agreement of the class boundaries predicted by SVM for $E=2$ kcal/mol (and $E=1$ kcal/mol) with the invariant manifolds that separate different dynamical behaviour.

\section{Results and discussion}
\subsection{Prediction using active learning}
\label{sec:active}

Under the assumption that integrating trajectories is more expensive than re-training SVM, we propose starting with a coarse initial grid and adaptively adding new points on the predicted boundary randomly sampled using a uniform distribution. As new points are added using successively accurate approximation of the boundary between the classes, this approach falls under active learning \cite{settles09,Kremer14}. This way it is possible to sample the surface more densely near boundaries between the classes and less densely elsewhere. By adding new points on the predicted boundary, SVM requires relatively few points to reach a qualitatively correct approximation of the class boundaries. 
  
For the sake of shorter computational time we adaptively sample 4 points at every iteration when calculating results shown in Fig. \ref{fig:adaptive_r35} and we expect more accurate results when the model is re-trained after every new sample point. We discard sample points outside of the energetically accessible area. The predicted boundary is obtained as a set of (unique) points that form the contours of an array of predicted class labels on a regular grid using scikit-image's \cite{scikit-image} find\_contours function. If required, the sampling may be biased toward a particular class by duplicating the corresponding contour line or sampling each contour line separately.

\begin{figure}
 \centering
 \includegraphics[width=0.49\textwidth]{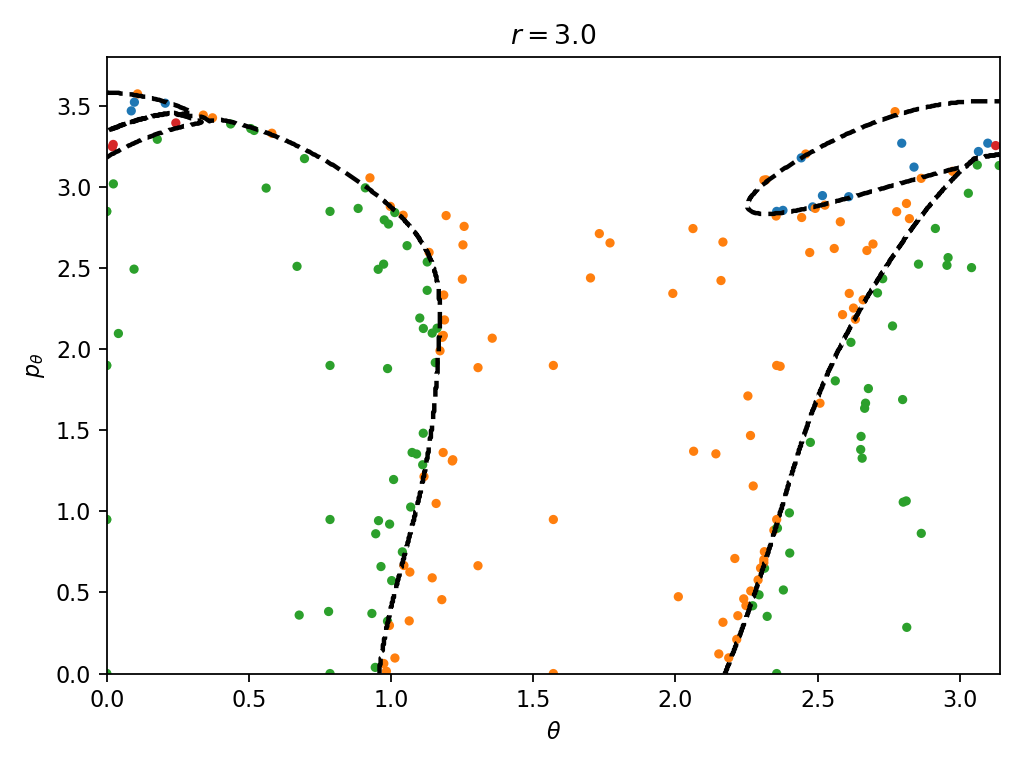}
 \includegraphics[width=0.49\textwidth]{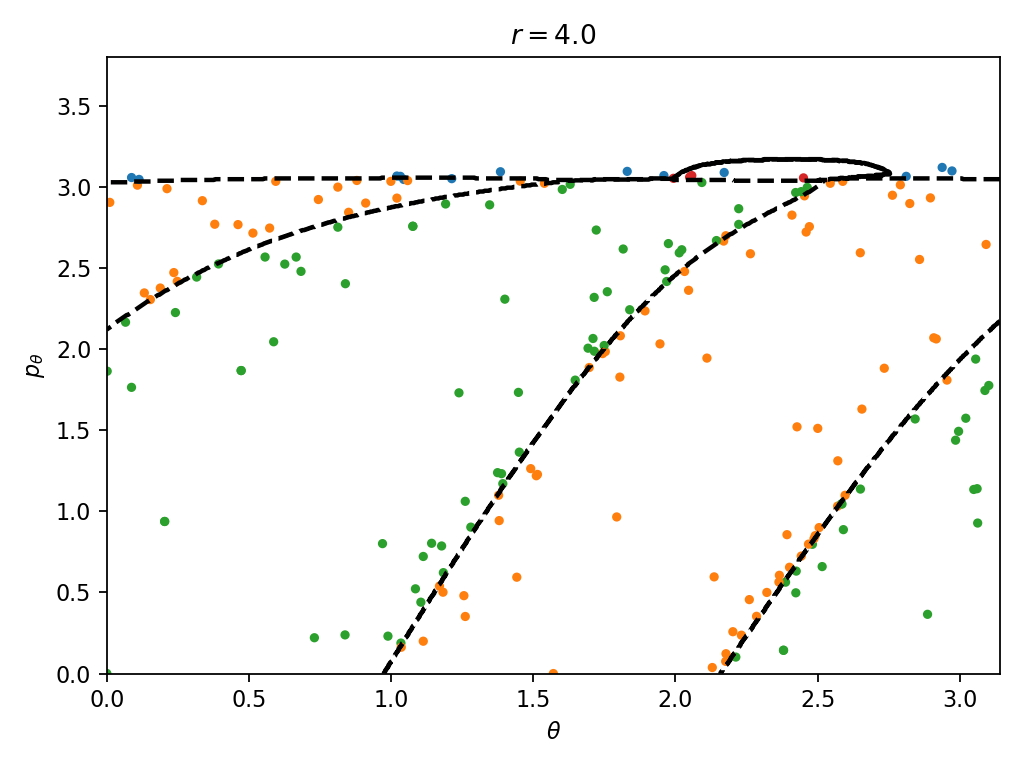}
 \includegraphics[width=0.49\textwidth]{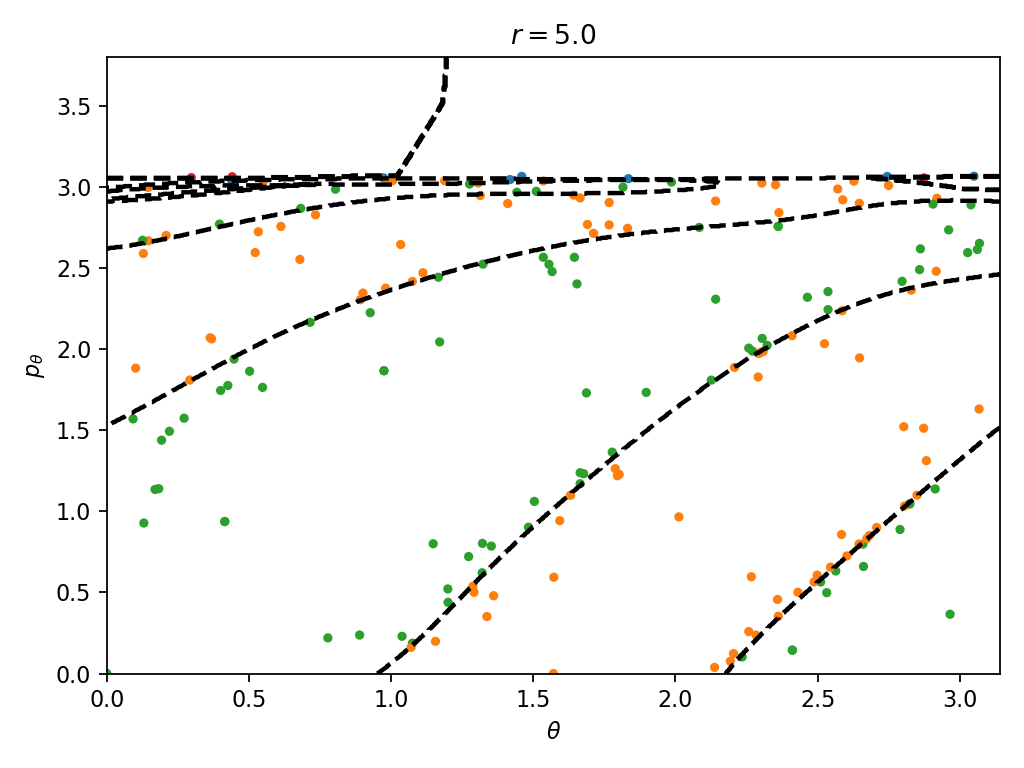}
 \includegraphics[width=0.49\textwidth]{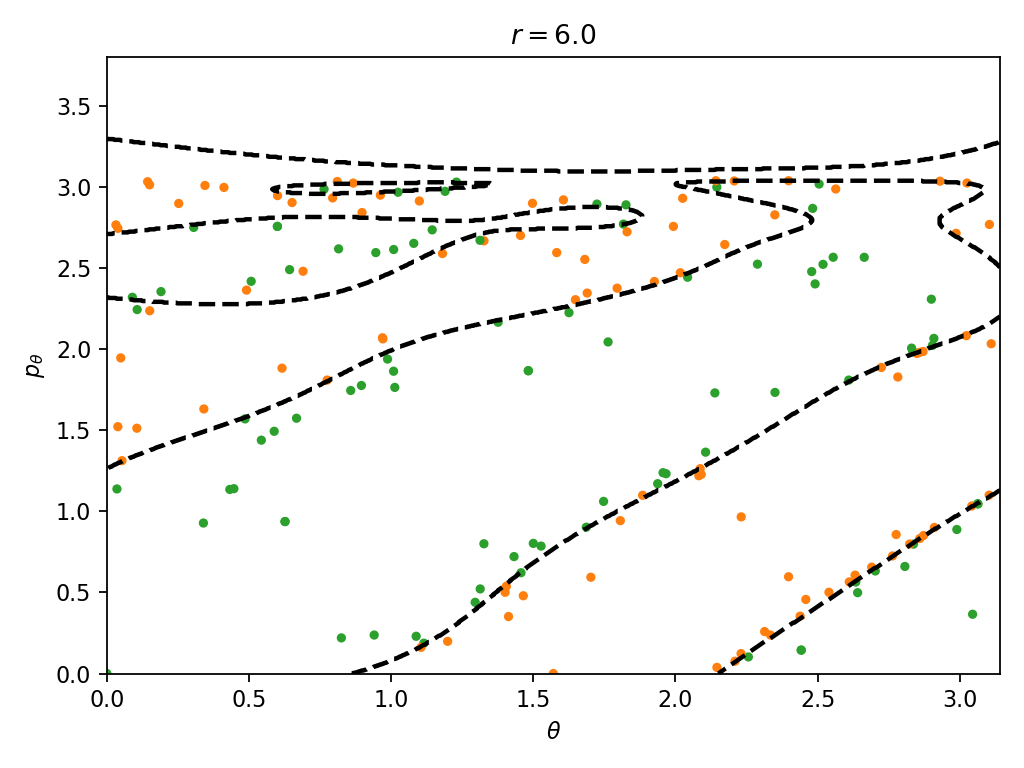} 
 \caption{Labelled initial conditions of trajectories on multiple surfaces $r=3, 4, 5, 6$ on an initial 5x5 grid and sampled adaptively using 200 points as described in text. The classes are class 1 (green), class 2 (red), class 3 (orange), class 4 (blue). Predicted class boundaries are shown in dashed black.}
 \label{fig:adaptive_multi}
\end{figure}

Classifying points on multiple surfaces simultaneously leads to a more accurate predictions and allows us to identify the most appropriate surface for further phase space analysis with minimal a priori knowledge  of the dynamics in the system. Fig. \ref{fig:adaptive_multi} shows the class boundaries predicted by SVM on the surfaces $r=3,4,5,6$ for an initial 5x5 grid and 4 adaptively sampled points per iteration until we reach a total of 200 points. At every iteration we add points on a different surface. Even though the initial data set is unbalanced, all but one of the initial 5x5 points correspond to class 1 (imminent dissociation and association) and class 3 (imminent dissociation only), the qualitatively correct structure of invariant manifolds is identified.
 
\begin{figure}
 \centering
 \includegraphics[width=\textwidth]{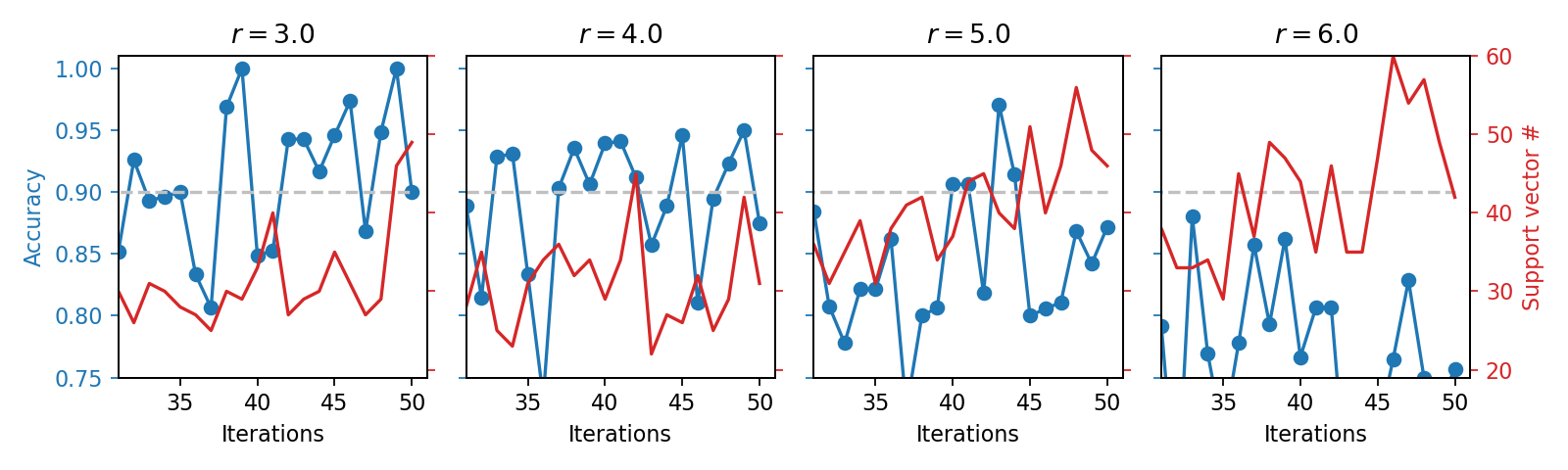}
 \caption{Plots comparing accuracy and number of support vectors for surfaces displayed in Fig. \ref{fig:adaptive_multi}. The more complicated geometry of class boundaries on $r=5$ and $r=6$ requires more support vectors and results in lower accuracy.}
 \label{fig:adaptive_accuracy}
\end{figure}

The most appropriate surface for a subsequent analysis is the surface with the simplest class boundaries, which is the one where SVM reaches the highest accuracy and requires the fewest support vectors as shown in Fig. \ref{fig:adaptive_accuracy}. This agrees with the geometric interpretation of SVM.
 
Despite the relatively high prediction accuracy on $r=3,4$ achieved for 200 points, it is evident from Fig. \ref{fig:adaptive_accuracy} that the results are not robust. Accuracy varies between $0.75$ and $1$ depending on the test/train split, which suffices for qualitative purposes, but may be limiting for quantitative investigations. For 500 points, such as the example shown in Fig. \ref{fig:adaptive_r35}, this fluctuation reduces to the range between $0.85$ and $1$. We note that the prediction accuracy can be improved by optimising the choice of surfaces, density of the initial grid and number of points added at every iteration.
 
On the other hand, a noticeably lower accuracy reached on $r=5,6$ and higher number of support vectors correspond to the increasingly distorted boundaries with $r$ approaching the centrifugal barrier slightly below $r=7$. Increasing the number of sample points to 500 does not improve accuracy significantly. We remark that none of the sampled point in class 2 and class 4 corresponds to a trajectory that reaches $r=6$ and all of them return to one of the surfaces $r=3,4,5$. Thus SVM attains a lower accuracy on $r=6$ despite the classification being binary. Consequently SVM not only provides the class boundaries and a decision function, but it can be used to evaluate how good/bad a given surface is to study the given class boundaries.

\subsection{Phase space structures}
\label{sec:structures}
Here we briefly discuss known phase space structures that are approximated by the class boundaries predicted by SVM. The class boundaries are formed by the intersection of the stable and unstable invariant manifolds of unstable periodic orbits with the surface. The invariant manifolds guide the trajectories from the potential well to dissociation or back to one of the potential wells \cite{mauguiere2014roaming,mauguiere2014multiple,krajnak2018phase}.

\begin{figure}
 \centering
  \includegraphics[height=0.4\textwidth]{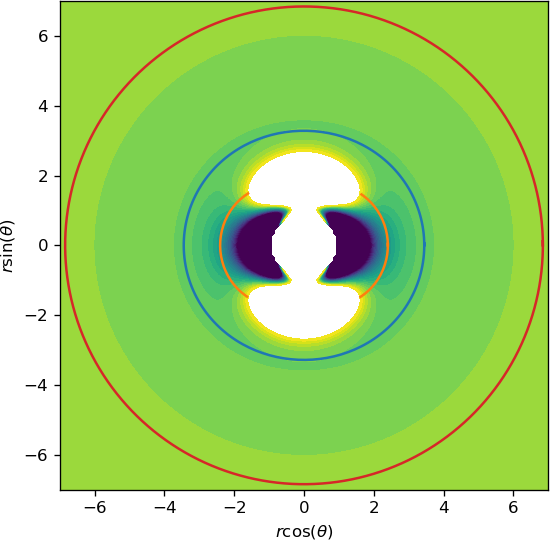}
  \caption{Configuration space projections of inner (orange), middle (blue) and outer (red) periodic orbits in Chesnavich's CH$_4^+$ system for the total energy $E=2$ kcal/mol plotted on top of a contour plot of potential energy $U$. Dividing surfaces project onto the projection of the corresponding orbits. These surfaces are used to define types of dynamical behaviour.}
 \label{fig:orbits}
\end{figure}

There are three important families of periodic orbits and at every fixed energy $0 \leq E \leq 2.5$ each family contains two orbits related by symmetry:
\begin{itemize}
 \item two inner periodic orbits (orange in Fig. \ref{fig:orbits}) delimit two potential wells corresponding to bound isomers of CH$_4^+$,
 \item two outer orbits (one clockwise, one counter-clockwise) due to a centrifugal barrier \cite{krajnak2018phase}, that delimit the dissociated states (red in Fig. \ref{fig:orbits}),
 \item two middle orbits (one clockwise, one counter-clockwise) in the flat region crucial for the definition of roaming.
\end{itemize}

Based on dynamical properties, previous works on Chesnavich's CH$_4^+$ system distinguished between the following types of dynamics: direct dissociation, roaming, isomerisation and non-reactive trajectories. These types are defined using dividing surfaces (DS) constructed from periodic orbits. We provide the construction details and define types of trajectories in Supplementary information. As found by \cite{krajnak2018phase, krajnak2018influence}, all types of trajectories cross the outward half of the middle DS, which is a surface that has the same configuration space projection as the middle orbits. The inner and outer DSs are defined analogously, see Supplementary information. The middle DS is hence the ideal surface to study the dynamics in Chesnavich's CH$_4^+$ system.

\begin{figure}
 \centering
  \includegraphics[height=0.5\textwidth]{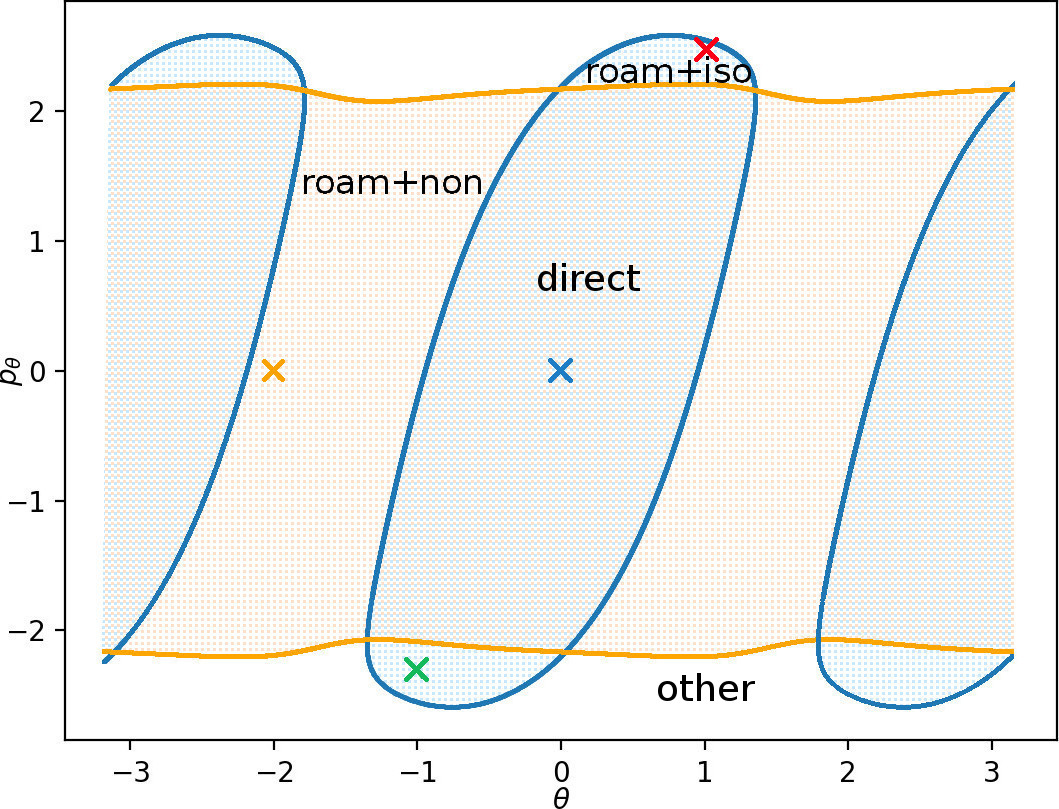}
  \caption{Reactive islands for $E=2$ kcal/mol formed by unstable invariant manifolds of the inner orbits (blue line) and stable invariant manifolds of the outer orbits (orange line) on the middle DS. These manifolds separate different classes of dynamics. Intersections of representative trajectories from Fig. \ref{fig:types} with the middle DS are indicated by crosses: direct dissociation (blue), roaming (red), isomerisation (green) and non-reactivity (yellow).}
 \label{fig:islands}
\end{figure}

Fig. \ref{fig:islands} shows the intersection of the middle DS with the unstable invariant manifolds of the inner orbits, guiding trajectories away from the inner DSs, and the stable invariant manifolds of the outer orbits conveying trajectories to the outer DS and into dissociated states. We would like to point out that the invariant manifolds intersect this surface infinitely many times, forming a fractal structure similar to that of Smale's horseshoe map. We only show the first intersection, that is the closest along the manifold to the corresponding periodic. In dynamical systems literature, these intersections are known as the first reactive islands \cite{DeVogelaere1955,PollakChild80,berne_isomerization_1982,de_leon_order_1989,marston_reactive_1989,OzoriodeAlmeida90}. Every subsequent reactive islands is an image of the first island under the return map for unstable manifolds or the pre-image under the return map for stable manifolds.

The first reactive islands allow us to identify the first and last intersections of the enclosed trajectories with the middle DS. It allows us to conclude whether a trajectory does immediately originate from a well or not, and whether a trajectory is going to dissociate immediately or not. Consequently we can distinguish between direct dissociation, roaming + isomerisation, roaming + non-reactivity, and other trajectories that remain in the region between the inner and outer DSs as shown in Fig. \ref{fig:islands}.

Images or pre-images of the first reactive islands can be calculated numerically using the return map. This approach of obtaining higher order reactive islands is preferable to directly using SVM, given that all higher order reactive islands in this system have fractal boundaries.

\subsection{Discovering phase space structures with SVM}
\label{sec:comparison}

As remarked in Sec. \ref{sec:surfaces}, the majority, but not all, of initial conditions in each class can be related to a dynamical behaviour induced by the intersection of stable and unstable invariant manifolds mentioned above. Initial conditions in Class 1 lead predominantly to direct dissociation, class 2 largely correspond to roaming + isomerisation, class 3 to roaming + non-reactivity and class 4 to other trajectories that remain between the inner and outer DSs.

\begin{figure}
 \centering
 \includegraphics[width=0.49\textwidth]{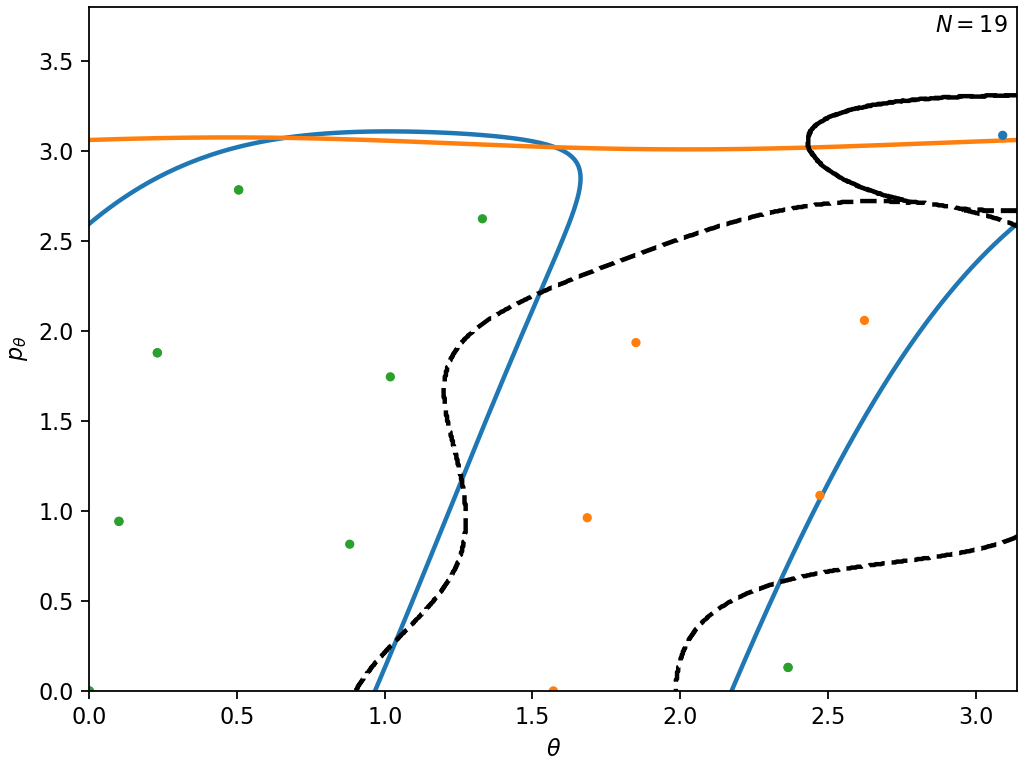}
 \includegraphics[width=0.49\textwidth]{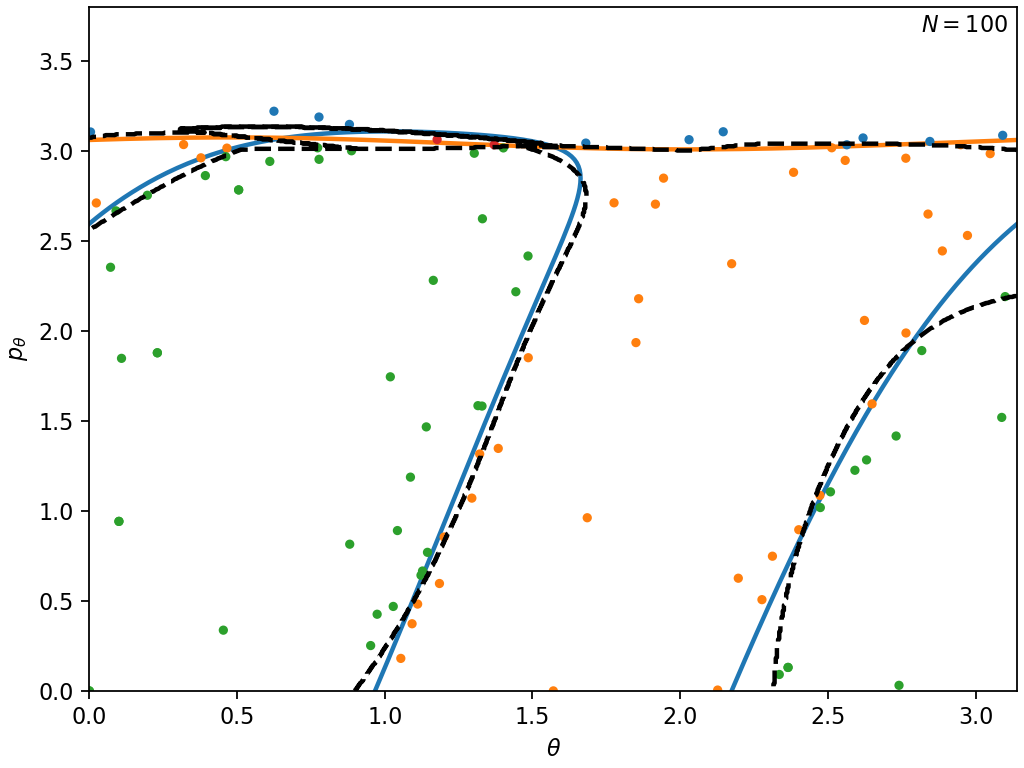}
 \includegraphics[width=0.49\textwidth]{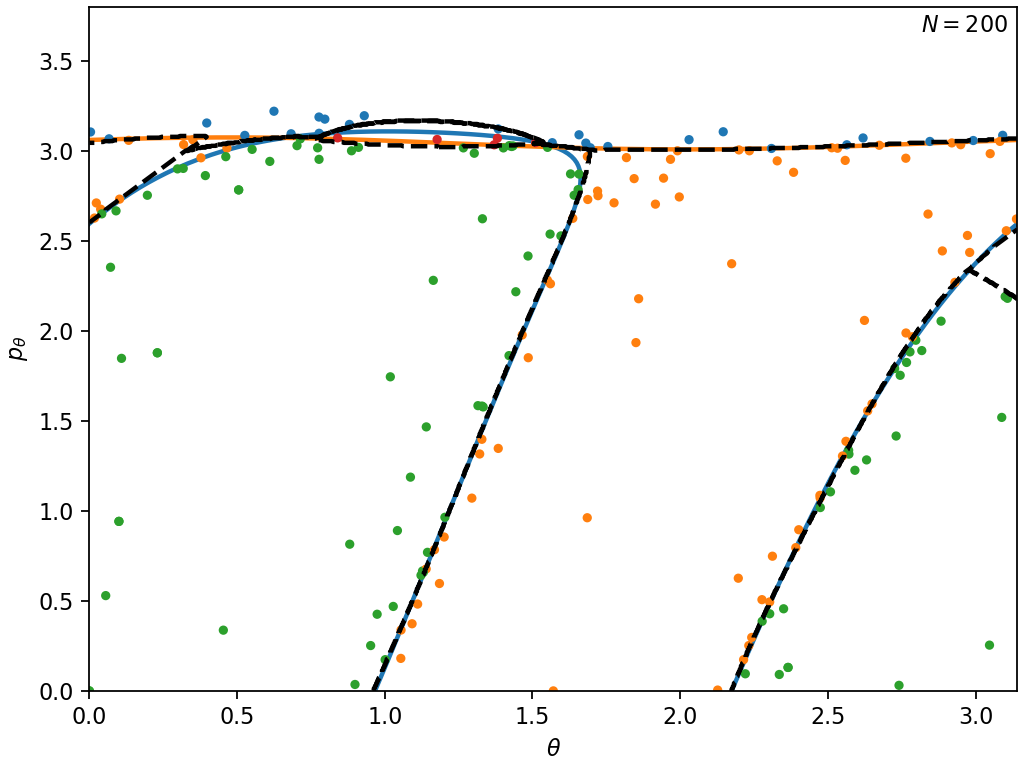}
 \includegraphics[width=0.49\textwidth]{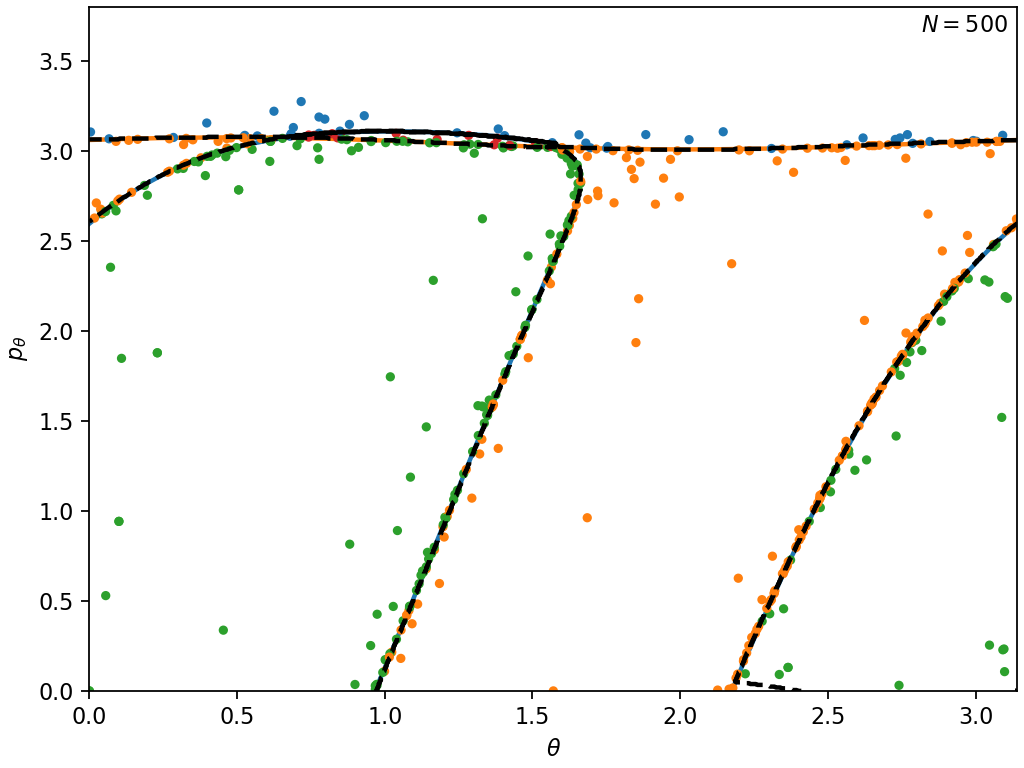}
 \caption{Convergence of SVM predictions and their agreement with known invariant manifolds on $r=3.5$ as predicted by SVM using $N=19,\ 100,\ 200,\ 500$ points, where $N=19$ is the initial 5x5 grid. The classes are class 1 (green), class 2 (red), class 3 (orange), class 4 (blue). Predicted class boundaries are shown in dashed black, overlayed with directly calculated invariant manifolds from Fig. \ref{fig:islands}. The procedure trained SVM on the surfaces $r=3,3.5,4$ simultaneously.}
 \label{fig:adaptive_r35}
 \end{figure}

The class boundaries identified by SVM ultimately approximate invariant manifolds identified in previous works and as discussed in Sec. \ref{sec:structures} to mediate the reaction dynamics in this system. Fig. \ref{fig:adaptive_r35} illustrates the convergence of SVM predictions and their agreement with known invariant manifolds, showing the initial 5x5 grid and subsequently at a total of 100, 200 and 500 sample points. Most notably, the intersection of invariant manifolds that causes roaming \cite{krajnak2018phase} is identified quickly. While SVM gives a clear indication on where roaming and isomerising trajectories are located, the precise distinction between roaming and isomerisation is not possible due to the fractal boundary separating them.

\section{Conclusions and Outlook}
\label{sec:concl}

In this paper we have shown how to develop a machine learning approach to discover the phase space structure that distinguishes between distinct reaction pathways using support vector machines (SVM). The approach uses minimal a priori knowledge of the dynamics in the system.  Given the definitions of classes of trajectories, we derive association and dissociation thresholds from properties of the potential energy surface. Our approach then classifies initial conditions on multiple surfaces using an active learning procedure based on SVM and selects the most appropriate surface to study the identified class boundaries. We show that the class boundaries approximate known phase space structures that have been computed in previous works on Chesnavich's model as the structures governing the dynamics in the model. The advantage of our SVM-based approach is that it requires calculating significantly fewer trajectories. This suggests that our approach may be successful in higher dimensional models of reaction dynamics, and this will be a focus for future work. In particular, the work on reactive islands in three degrees-of-freedom Hamiltonian systems described in \cite{krajnak2021reactive} could serve as a benchmark.

\section*{Acknowledgments}
\noindent We acknowledge the support of  EPSRC Grant no. EP/P021123/1 and Office of Naval Research (Grant No.~N000141712220).

% \appendix

\newpage
\bibliography{ri_3dof,roaming_v8,SNreac}

\section*{Supplementary information}

\subsection*{Support vector machine for classifying trajectories}

The classification algorithms for support vector machines (SVM)~\cite{Cortes95,Vapnik96,Vapnik2000} construct a nonlinear boundary between different classes by lifting into a high dimensional space and constructing a hyperplane that separates the classes. In our case, the classes correspond to trajectories with qualitatively distinct dynamics and initialized on the section $r = const$ and $\dot{r} > 0$. For Chesnavich's model, by qualitatively distinct dynamics we mean trajectories that lead to direct dissociation, roaming, isomerisation, or non-reactive (as shown in Fig.~\ref{fig:types}). As we discuss in Section~\ref{sec:structures}, the nonlinear boundaries that classify the initial conditions are obtained from the phase space analysis. To be precise, we show that the boundaries between initial conditions are given by the intersection of stable and unstable invariant manifolds of unstable periodic orbits with the chosen section. Therefore, a SVM classification algorithm, also referred to as support vector classifier (SVC), can approximate these manifolds given appropriate training data.

\begin{figure}[!h]
	\centering
	\includegraphics[width=0.75\textwidth]{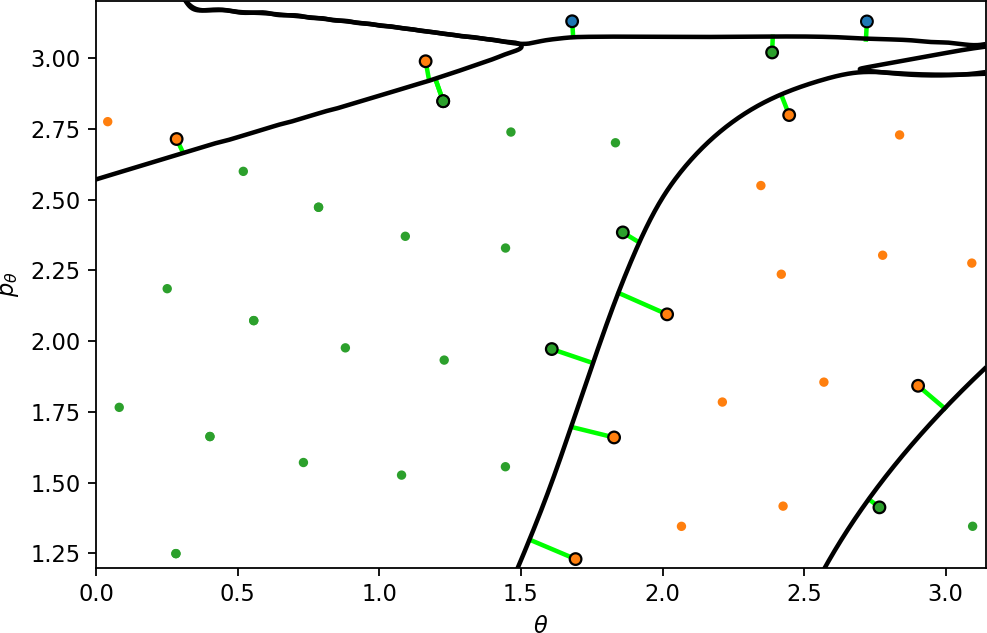}
	\caption{An illustration of decision boundaries (shown as black curves) between three classes of data (blue, orange and green) calculated by the support vector classifier with radial basis function kernel. The distance between the boundary and the closest points $P_i$ of every class, in this case the support vectors, is highlighted in light green. In the case of active learning, we show that adding new training data on these decision boundaries leads to faster learning by the support vector classifier.}
	\label{fig:svm_illustration}
\end{figure}

Following \cite{pozun2012optimizing}, we use the \verb|scikit-learn| \cite{scikit-learn}~\cite{Chang2011} library's implementation of support vector classifier. The implementation can be used with various kernels, of which the radial basis function kernel is best suited to approximate nonlinear boundaries, which are topological circles as discussed in section~\ref{sec:structures}, between the different classes of trajectories. After learning the kernel (or optimization problem to find the hyperparameters) using the training data, a previously unseen initial condition $P$ is predicted to belong to a class using the decision function
\begin{equation}
	\sum\limits_{i=1}^N \alpha_i l_i e^{-\gamma ||P_i-P||^2},
\end{equation}
where $\gamma>0$ controls the width of the Gaussian, $l_i=\pm1$ are class labels of training data $P_i$, $C\geq\alpha_i\geq0$ are weights calculated by the algorithm, of which only a number of the weights $\alpha_i$ is non-zero, and $N$ is the number of initial conditions on the two-dimensional section. These weights correspond to the subset of the training data $P_i$ called support vectors. The weights $\alpha_i$ are calculated by SVC such that the distance between the predicted boundary and the closest points $P_i$ of every class is maximised, as illustrated in Fig.~\ref{fig:svm_illustration}.

The upper bound $C$ on weights $\alpha_i$ is a user defined value that controls the complexity of the decision boundary - a low value of $C$ gives a smoother decision boundary, while a high value of $C$ leads to higher accuracy. In this article, we first perform a search over a wide interval of $C,\gamma$ values as shown in~\cite{naik_support_2021}. Then, a smaller interval for both parameters is chosen for each of the support vector classifier approaches. The cross-validation ensures that the trained model does not suffer from over-fitting by splitting the training data into 5 folds, each of which is used as a test set with the remaining four as training set.

\subsection*{Construction of dividing surfaces and types of trajectories}

In systems with 2 DoF, dividing surfaces (DS) are constructued using periodic orbits as follows.
For every orbit $\Gamma$ with the configuration space projection $(r_\Gamma, \theta_\Gamma)$, the corresponding DS is defined as the set of all points, $(r_\Gamma, p_r, \theta_\Gamma, p_\theta)$, on the energy surface such that $H(r_\Gamma, p_r, \theta_\Gamma, p_\theta)=E$. The inner DSs associated with the inner periodic orbits are known to be spheres, one corresponding to each periodic orbit. In contrast, the middle and outer DSs associated with the middle and outer periodic orbits are known to be tori, each containing two orbits. Because a trajectory can intersect a DS at a given point in two different directions, for example $\dot{r}>0$ and $\dot{r}<0$ in case of the outer DS.

In all of the following when we refer to a DS we mean its outward half.
Directly dissociating trajectories evolve from a well to the dissociated states passing through an inner, the middle and the outer DS. Isomerising trajectories originate in a well, pass through the middle DS and end in a well without ever reaching the outer DS. Similarly non-reactive trajectories go from dissociated states to dissociated states via the middle DS without ever reaching an inner DS. Roaming is defined as the evolution of trajectories from either of the potential wells to the dissociated states while crossing the middle DS more than once \cite{mauguiere2014roaming}.

\subsection*{Comment on fractal structures}
% \label{sec:fractals}

For some systems, higher order reactive islands can be determined in the same fashion as the first ones. As a consequence of heteroclinic intersections, higher order reactive islands in this system have a fractal structure, which is not suited for SVM due to the following reasons:
\begin{itemize}
 \item Depending on sampling density, a finite amount of data resolves fractal structures only up to a limited resolution.
 \item The heteroclinic intersections result in parts of a reactive island to be deformed into narrowing bands. Radial basis functions underlying our SVM approach have a fixed radius which are not suitable to approximate structures with varying ``widths''. While it will perform well in situations when the boundaries/manifolds between two classes are sufficiently far apart, the proximity of multiple boundaries/manifolds leads to inaccuracies.
\end{itemize}
\end{document}